\documentclass[sigconf]{acmart}
\usepackage{algpseudocode}
\usepackage{listings}

\usepackage{bm}
\usepackage[ruled,vlined,linesnumbered]{algorithm2e}
\usepackage{subfigure}
\usepackage{xcolor}
\usepackage{graphicx}
\usepackage{multirow}
\usepackage{pifont}
\newcommand{\cmark}{\ding{51}} 
\newcommand{\xmark}{\ding{55}} 

\usepackage{amsmath,amsfonts}

\usepackage{booktabs} 
\usepackage{tabularx} 
\usepackage{graphicx} 
\usepackage{subcaption} 

\AtBeginDocument{%
  }

\copyrightyear{2025}
\acmYear{2025}
\setcopyright{acmlicensed}\acmConference[MM '25]{Proceedings of the 33rd ACM International Conference on Multimedia}{October 27--31, 2025}{Dublin, Ireland}
\acmBooktitle{Proceedings of the 33rd ACM International Conference on Multimedia (MM '25), October 27--31, 2025, Dublin, Ireland}
\acmDOI{10.1145/3746027.3761997}
\acmISBN{979-8-4007-2035-2/2025/10}

\begin{document}

\title{Hierarchical Vision-Language Reasoning for Multimodal Multiple-Choice Question Answering}


\author{Ao Zhou}
\authornote{Equal contribution.}
\affiliation{%
  \institution{State Key Laboratory for Novel Software Technology, Nanjing University}
  \city{Nanjing}
  \country{China}
}\email{zacqupt@gmail.com}

\author{Zebo Gu}
\authornotemark[1]
\affiliation{%
  \institution{Chongqing University of Posts and Telecommunications}
  \city{Chongqing}
  \country{China}}
\email{s240231130@stu.cqupt.edu.cn}

\author{Tenghao Sun}
\authornotemark[1]
\affiliation{%
  \institution{Chongqing University of Posts and Telecommunications}
  \city{Chongqing}
  \country{China}}
\email{kudohao@163.com}

\author{Jiawen Chen}
\authornotemark[1]  
\affiliation{%
  \institution{Chongqing University of Posts and Telecommunications}
  \city{Chongqing}
  \country{China}}
\email{s240233002@stu.cqupt.edu.cn}
\author{Mingsheng Tu}

\affiliation{%
  \institution{Chongqing University of Posts and Telecommunications}
  \city{Chongqing}
  \country{China}}
\email{tumingsheng25@gmail.com}

\author{Zifeng Cheng}
\affiliation{%
  \institution{State Key Laboratory for Novel Software Technology, Nanjing University
}
  \city{Nanjing}
  \country{China}
}
\email{chengzf@smail.nju.edu.cn}

\author{Yafeng Yin}
\affiliation{%
  \institution{State Key Laboratory for Novel Software Technology, Nanjing University}
  \city{Nanjing}
  \country{China}
}
\email{yafeng@nju.edu.cn}

\author{Zhiwei Jiang}
\authornote{Corresponding author.}
\affiliation{%
  \institution{State Key Laboratory for Novel Software Technology, Nanjing University
}
  \city{Nanjing}
  \country{China}}
\email{jzw@nju.edu.cn}

\author{Qing Gu}
\affiliation{%
  \institution{State Key Laboratory for Novel Software Technology, Nanjing University}
  \city{Nanjing}
  \country{China}}
\email{guq@nju.edu.cn}

\renewcommand{\shortauthors}{Ao Zhou et al.}

\begin{abstract}
Multimodal Large Language Models (MLLMs) have demonstrated remarkable multimodal understanding capabilities in Visual Question Answering (VQA) tasks by integrating visual and textual features. However, under the challenging ten-choice question evaluation paradigm, existing methods still exhibit significant limitations when processing PDF documents with complex layouts and lengthy content. Notably, current mainstream models suffer from a strong bias toward English training data, resulting in suboptimal performance for Japanese and other language scenarios. To address these challenges, this paper proposes a novel Japanese PDF document understanding framework that combines multimodal hierarchical reasoning mechanisms with Colqwen-optimized retrieval methods, while innovatively introducing a semantic verification strategy through sub-question decomposition. Experimental results demonstrate that our framework not only significantly enhances the model's deep semantic parsing capability for complex documents, but also exhibits superior robustness in practical application scenarios.
\end{abstract}




\maketitle

\section{Introduction}
The rapid advancement of Large Language Models (LLMs) has significantly expanded the frontiers of Natural Language Processing (NLP) technologies, demonstrating unprecedented potential in both daily life and commercial applications. In this evolutionary process, Multimodal Large Language Models (MLLMs) \cite{mllm_survey,mllm_survey2} and Vision-Language Models (VLMs) \cite{vlm_survey} have established new technical paradigms for handling complex cross-modal tasks, such as Visual Question Answering (VQA), through their integrated visual-textual comprehension capabilities.
However, existing MLLMs still face significant challenges when processing QA tasks based on lengthy PDF documents, where the unique structural complexity of such documents presents a critical barrier to commercial deployment. The primary challenges stem from both the extensive document length (often spanning dozens to hundreds of pages) and the complex non-linear layouts composed of diverse visual elements including tables, charts, and diagrams. 
The current heavy reliance on English-language data resources creates a pronounced gap for non-English languages \cite{VQAV1,VQAV2}, with Japanese document-level VQA being particularly underserved as \textit{JDocQA} remains the sole publicly available benchmark.

Traditional Optical Character Recognition (OCR) techniques and pure-text language models struggle to effectively parse such structured content \cite{Mplug-paperowl}, while comprehensive document understanding \cite{Document_understanding} often requires establishing semantic correlations among elements. To overcome these limitations, the research community has developed specialized VLMs for document understanding, such as mPLUG-DocOwl \cite{Mplug-docowl}, Docopilot \cite{Docopilot}. Meanwhile, general-purpose MLLMs, including the GPT-4 \cite{Gpt-4}, GLM4.1 \cite{GLM-4.1}, as well as Qwen-VL \cite{Qwenvl} and Kimi-VL \cite{Kimi}, have also demonstrated increasing adaptability in document image processing.

This study conducts an evaluation on a newly established Japanese PDF document understanding benchmark. The benchmark is constructed from the JDocQA corpus as a domain-specific question–answering dataset, where the task is to select the correct answer from ten candidates. We propose an innovative framework that integrates multimodal information processing with a hierarchical reasoning mechanism. 
Specifically, ColQwen \cite{ColPali} is employed to identify the top-3 most semantically relevant pages from the entire PDF document based on queries. 
Subsequently, leveraging a large language model with carefully designed prompts, the original question is decomposed into a sequence of sub-questions, and a semantic verification mechanism is introduced to mitigate semantic drift, thereby significantly improving the accuracy and robustness of retrieval results. Furthermore, prompt engineering is utilized to produce a structured output format that preserves the complete reasoning trace.
We perform zero-shot evaluation of multiple state-of-the-art MLLMs using our proposed framework on the Japanese PDF document understanding benchmark, systematically assessing their comprehension capabilities without task-specific fine-tuning.
The evaluation aims to compare the models’ effectiveness in accurately selecting the correct answer from ten candidates without task-specific fine-tuning, thereby providing a fair assessment of their generalization ability. Performance is measured using the Public Score metric, enabling consistent comparison across different model architectures and scales.

\section{Related Work}

\subsection{Multimodal Large Language Models}
Early Multimodal Large Language Models (MLLMs) usually use Large Language Models (LLMs) as controllers to parse prompts, break down complex questions, and call visual foundation models (VFMs) like CLIP \cite{CLIP}. For example, Visual ChatGPT combines VFMs such as BLIP \cite{Blip} and Stable Diffusion \cite{stable_diffu} through a prompt manager, enabling step-by-step reasoning and accurate image generation. Another approach converts images into descriptive text to help reasoning. IMG2LLM \cite{IMG2LLM} improves this by focusing on image regions related to the question to generate better captions. Although these methods let pure-text LLMs handle multimodal tasks, they do not truly understand images, relying instead on external visual models or image-to-text conversion. To solve this, recent research \cite{Qwenvl,GLM-4.1}aims to build general-purpose multimodal LLMs that directly understand images by integrating multiple modalities into one unified model.

The typical training paradigm for general MLLMs consists of two stages. In the first stage, the LLM is frozen while only the visual encoder and image-to-text alignment modules are trained to map visual features into the LLM’s textual space, as seen in the Q-Former of BLIP-2 \cite{Blip-2}. In the second stage, the visual modules are fixed, and the LLM is fine-tuned on multimodal data using parameter-efficient methods such as LoRA \cite{lora} or Q-LoRA \cite{Qlora}, for example, in LLaVA \cite{Lava}. The main challenge in MLLMs lies in designing effective image-to-text alignment modules. Different MLLMs adopt various alignment techniques; for instance, Flamingo \cite{Flamingo} uses a visual encoder combined with a perceiver resampler to generate fixed-length feature sequences, which are then integrated with cross-attention layers to enhance alignment between visual and textual modalities. Among commercial models, GPT-4v~\cite{Gpt-4} enables seamless interleaved image–text processing, and Gemini Pro~\cite{Gemini} delivers strong performance across multimodal tasks. In the open-source domain, representative models such as Qwen~\cite{Qwenvl,Qwenvl2}, Kimi~\cite{Kimi}, InternVL-3~\cite{Internvl}, and mPLUG-OWL~\cite{Mplug-docowl,Mplug-paperowl} have made significant progress in vision–language tasks and complex reasoning.

\subsection{Retrieval-Augmented Generation}
Retrieval-Augmented Generation (RAG) \cite{RAG_survey} is a framework that combines information retrieval and generative models to enhance the performance of LLMs by leveraging external knowledge. The RAG framework mainly consists of three components: retrieval, generation, and augmentation. 
The retrieval module utilizes either sparse retrieval methods such as BM25 \cite{BM25} or dense retrieval techniques like Dense Passage Retriever (DPR) \cite{DPR} to find relevant documents from large corpora. The generation module produces answers based on the retrieved content, commonly employing models such as BART \cite{BART} and T5 \cite{T5}. The augmentation component integrates the retrieved information into the generation process through various strategies, including input-layer concatenation (e.g., In-Context RALM \cite{RALM}), output-layer interpolation (e.g., kNN-LM \cite{Knn-LM}), and intermediate-layer interaction (e.g., RETRO \cite{RETRO}). This approach effectively improves the knowledge coverage and reasoning ability of LLMs, supporting zero-shot and few-shot tasks, and is widely applied in open-domain question answering, dialogue systems, and knowledge-intensive applications. 
For example, Atlas \cite{Atlas} and REALM \cite{REALM} leverage retrieval to achieve more accurate knowledge access, while multimodal models such as Flamingo \cite{Flamingo} and BLIP \cite{Blip} adopt retrieval augmentation to enhance visual-language understanding.

\subsection{Visual Question Answering}
Visual Question Answering (VQA) \cite{VQA_survey} represents a pivotal multimodal task situated at the confluence of computer vision and natural language processing, where the objective is to generate an answer \( A \) to a question \( Q \) given visual input \( V \), typically expressed as \( A = f(Q, V) \). Early VQA efforts, grounded in datasets such as VQA v1 \cite{VQAV1} based on MS COCO images, leveraged convolutional neural networks (e.g., VGG-Net \cite{VGG-Net}, Faster-RCNN \cite{Faster-r-cnn}) for visual feature extraction alongside recurrent models like LSTMs and GRUs \cite{VQA-1,VQA-2} for textual encoding. 
The introduction of attention mechanisms, including stacked attention and co-attention, marked a significant milestone by enabling models to selectively focus on relevant image regions conditioned on the question, thereby enhancing cross-modal feature fusion. Further, graph neural networks (GNNs) \cite{VQA_survey} have been employed to model complex relational structures between visual and textual elements via multimodal heterogeneous graphs. The advent of Transformer architectures revolutionized VQA, with pre-trained visual-language models such as BLIP utilizing self-attention and cross-attention to achieve robust multimodal integration and remarkable zero-shot performance. More recently, MLLMs like Flamingo \cite{Flamingo}, BLIP-2 \cite{Blip-2}, and InternVL \cite{Internvl}have demonstrated superior capabilities, particularly in open-ended and zero-shot scenarios. Beyond direct perception, contemporary approaches emphasize knowledge-based reasoning by structuring internal knowledge into triplets or incorporating external knowledge bases and passage retrieval to form enriched image-question-knowledge representations for multi-hop reasoning. Additionally, sophisticated reasoning techniques such as instruction tuning and Chain-of-Thought (COT) \cite{COT} prompting have been integrated into MLLMs to further improve answer accuracy. 

\begin{figure*}[!h]
\centering
\includegraphics[width=0.9\textwidth]{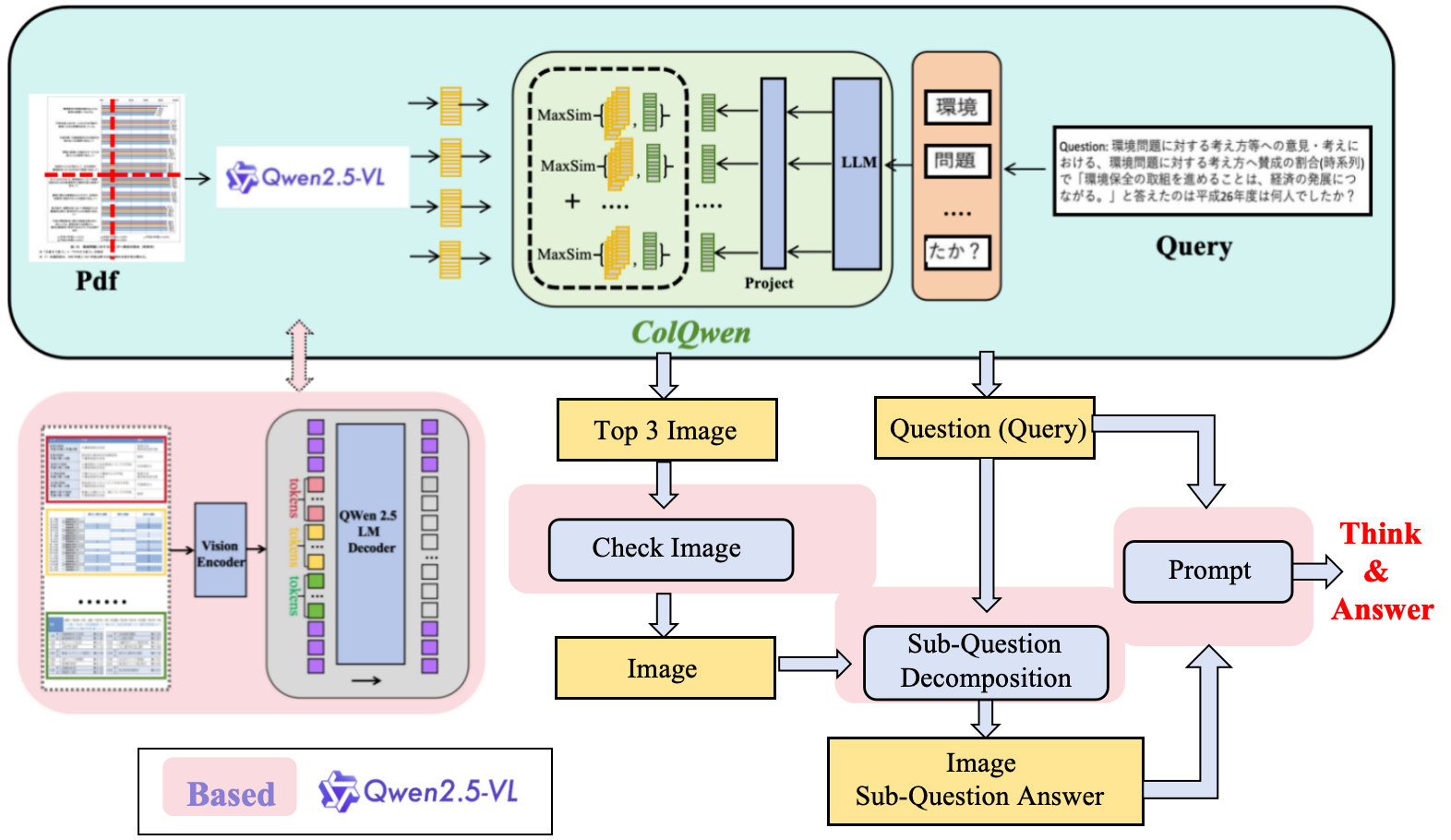} 
\caption{The workflow of proposed framework.}
\label{fig:framework}
\end{figure*}

\section{Proposed Method}

\subsection{Document Retrieval with Vision Language Models}
ColPali~\cite{ColPali} processes document page screenshots through a VLM pipeline. The input image is split into patches, encoded by a ViT, and the resulting patch embeddings are treated as soft tokens for the language model component (Gemma-2B in PaliGemma-3B). These embeddings are projected into a 128-dimensional space to form compact multi-vector representations. During retrieval, queries are encoded by the same Gemma-2B model, with similarity computed via a ColBERT-style late interaction mechanism:
\begin{equation}
\text{LI}(q, d) = \sum_{i=1}^{N_q} \max_{j \in [1, N_d]} \langle \mathbf{E}_q^{(i)}, \mathbf{E}_d^{(j)} \rangle
\end{equation}
where $\mathbf{E}_q \in \mathbb{R}^{N_q \times D}$ and $\mathbf{E}_d \in \mathbb{R}^{N_d \times D}$ are the query and document embedding matrices, respectively, $N_q$ and $N_d$ denote the number of tokens in the query and document, $\langle \cdot, \cdot \rangle$ is the inner product. The model is trained with in-batch negative sampling using a softplus contrastive loss:
\begin{equation}
\mathcal{L} = \frac{1}{b} \sum_{k=1}^{b} \log \Big( 1 + \exp\big( \max_{\substack{l=1 \ l \neq k}}^b \text{LI}(q_k, d_l) - \text{LI}(q_k, d_k) \big) \Big)
\end{equation}
where $b$ is the batch size, $q_k$ is the $k$-th query, $d_k$ is its corresponding positive document, and ${d_l}_{l \neq k}$ are in-batch negatives.

ColQwen2 builds upon ColPali with notable improvements in both architecture and training strategy. It supports dynamic image resolutions while preserving aspect ratios, enhancing robustness to diverse document layouts. The model is trained on a mix of public datasets and synthetic question-answer pairs generated by VLMs, with a focus on English to support strong zero-shot generalization across languages. ColQwen2 leverages efficient training techniques such as LoRA \cite{lora} and 8-bit optimizers to reduce memory and computation costs. It maintains the ColBERT-style \cite{Colbert} multi-vector late interaction mechanism, enabling fine-grained matching between queries and document patches.

At this stage, the PDF document is first split into individual pages, each of which is converted into a PNG image. Given a query, the ColQwen2 model is employed to retrieve the top three page images ranked by similarity scores. Subsequently, the QwenVL2.5 model is applied to these three retrieved images to further filter out irrelevant pages. This filtering process retains one to three relevant images, ensuring that at least one image is preserved for subsequent analysis.

\subsection{Sub Question Decomposition}
This module takes as input an image (e.g., statistical chart, table, or diagram), a question, and reference materials, and analyzes them to automatically generate multiple sub-questions necessary to solve the original question. First, it combines image recognition with text extraction to obtain structured data from the visual and textual content. The question is then semantically decomposed into logical steps, which are formulated as sub-questions. These sub-questions cover multi-step reasoning processes such as locating relevant data, extracting numerical values, identifying trends, and performing comparative analysis. While each sub-question is assigned a consistent answer during processing, the module’s final output is the single, conclusive answer to the original question. This design ensures transparency in the reasoning process while enabling efficient and accurate answer generation for complex problems.
\begin{figure*}[!h]
\centering
\begin{subfigure}[First step]{
  \includegraphics[width=0.57\textwidth]{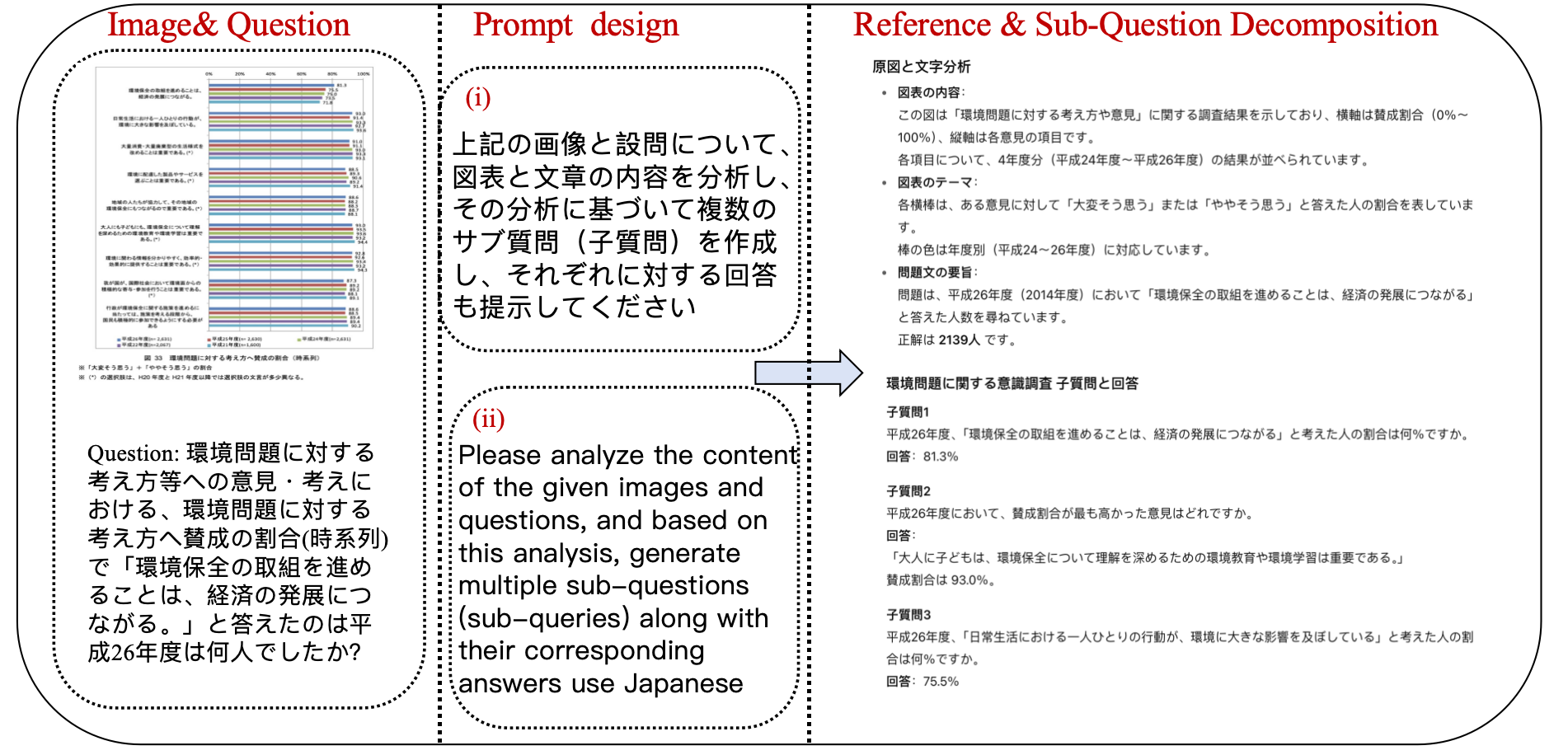}
  \label{fig:sub-question-decomp1}
  }
\end{subfigure}
\hfill
\begin{subfigure}[Second step]{
  \includegraphics[width=0.4\textwidth]{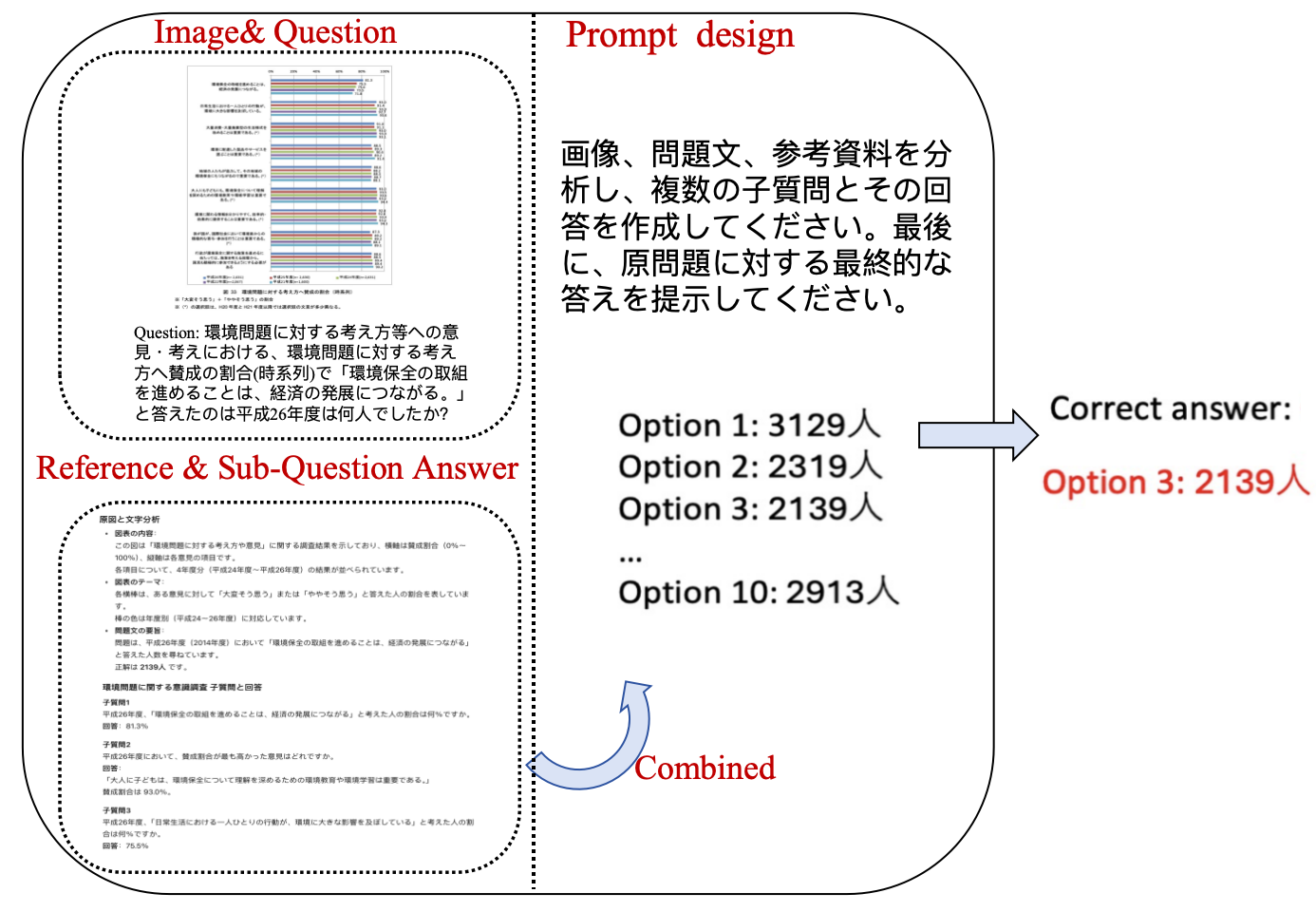}
  \label{fig:sub-question-decomp2}
  }
\end{subfigure}
\caption{Sub-question decomposition process.}
\label{fig:Sub-Question-Decomp}
\end{figure*}
As shown in Figure~\ref{fig:Sub-Question-Decomp}, the input chart presents survey results on ``perceptions and attitudes toward environmental issues,'' 
where the horizontal axis indicates the proportion of agreement (0\%--100\%), and the vertical axis lists different viewpoints. 
Each viewpoint contains comparative data for three fiscal years: Heisei 24 (2012), Heisei 25 (2013), and Heisei 26 (2014). 
Through OCR and structured data parsing, the model identifies the data types (e.g., percentages, year-wise comparisons) and associates them with their corresponding viewpoints.

When parsing the question text, the model recognizes that the original question is: 
``In the survey "Opinions and Perspectives on Environmental Issues", what was the number of respondents in fiscal year Heisei 26 (2014) who agreed with the statement "Environmental conservation efforts will lead to economic development" in the time-series data showing approval rates for environmental perspectives ?'' 
To enhance reasoning transparency and interpretability, as illustrated in Figure~\ref{fig:sub-question-decomp1}, 
we first propose the following prompt: 
\begin{itemize}
\item \textbf{First step prompt:} Given the above image and question, analyze the chart and textual content, generate multiple sub-questions based on the analysis, and provide answers to each.
\end{itemize}
When combined with Qwen2.5-VL, this prompt enables the model to decompose the original question into several logically related sub-questions and generate corresponding answers. 
For example:
\begin{itemize}
    \item \textbf{Sub-question:} What was the proportion of agreement with this viewpoint in fiscal year 2014? \\ \textbf{Answer:} 81.3\%.
    \item \textbf{Sub-question:} Which viewpoint had the highest proportion of agreement in fiscal year 2014? \\ \textbf{Answer:} Environmental education: 93.0\% .
\end{itemize}
These sub-questions encompass multiple reasoning types, including locating key information, comparing different viewpoints, and identifying trends. During this process, the model may also retrieve reference information. 
For instance:
\begin{itemize}
\item \textbf{Reference:} the survey chart's horizontal axis represents percentages, while the vertical axis lists different viewpoints or attitudes, each with data from fiscal year 2012 to 2024. \\
\item \textbf{Reference:} Each horizontal bar indicates the proportion of respondents who agreed (combining ``strongly agree'' and ``somewhat agree''). \\
\end{itemize}

As illustrated in Figure~\ref{fig:sub-question-decomp2}, building upon the explicit generation of intermediate reasoning steps (including sub-questions and their solutions) with reference materials serving as knowledge anchors, we further integrate multimodal information comprising: visual data (the original image), semantic context (the initial question), structured reasoning processes (decomposed sub-questions with verified answers), and external knowledge (reference materials). 
\begin{itemize}
    \item \textbf{Second step prompt:} Analyze the image, the question sentence, and the reference materials, create multiple sub-questions with their answers, and finally, provide the final answer to the original question.
\end{itemize}

Through this workflow, the final answer to the original question is obtained .

\subsection{Prompt Design}
In the first step, the prompts in the first phase were provided in both English and Japanese. Our investigation revealed that most existing MLLMs perform better on English corpora, while Japanese prompts maintain consistency when processing Japanese charts. Using both English and Japanese versions of the prompts helps cover a wider range of semantic expressions and reduces the risk of task failure caused by language comprehension bias. In the second-phase prompts, we used only the Japanese version. 
\begin{figure}[!h]
    \centering
    \includegraphics[width=0.9\linewidth]{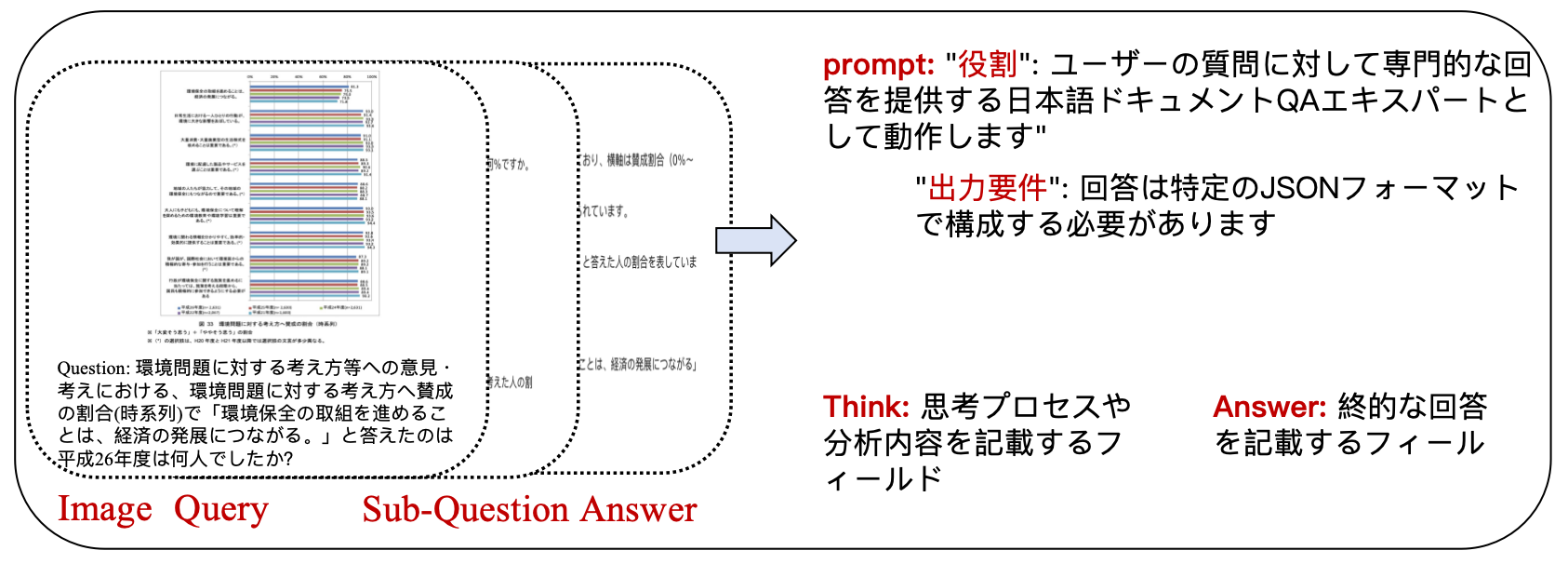}
    \caption{Prompt design in zero shot infer.}
    \label{fig:prompt}
\end{figure}
This is because the reference materials and sub-question--answer pairs generated in the first phase, which consist entirely of Japanese text, are used as the input for the second phase. 
By retaining only the Japanese prompts, the reasoning process remains consistent with the input language, which helps reduce information loss and ambiguity caused by cross-language conversion.

As shown in Figure~\ref{fig:prompt}, our second-phase research introduced an improved Japanese prompt engineering.
We formally define the model's operational role as a Japanese Document QA Specialist, implemented through the following structured prompt:
\begin{itemize}
\item  \textbf{First:} You are a Japanese Document QA Specialist.
\item  \textbf{Second:} Give structured JSON outputs with mandatory \texttt{think} (reasoning process) and \texttt{answer} (final answer) fields. 
\end{itemize}
This structure improves reasoning traceability and facilitates system integration for automated evaluation, answer provenance tracking, and knowledge base synchronization. 

\subsection{Additional Modules}
Specifically, we can introduce auxiliary modules to pre-lock relevant areas for retrieval.
\begin{figure}
    \centering
    \includegraphics[width=\linewidth]{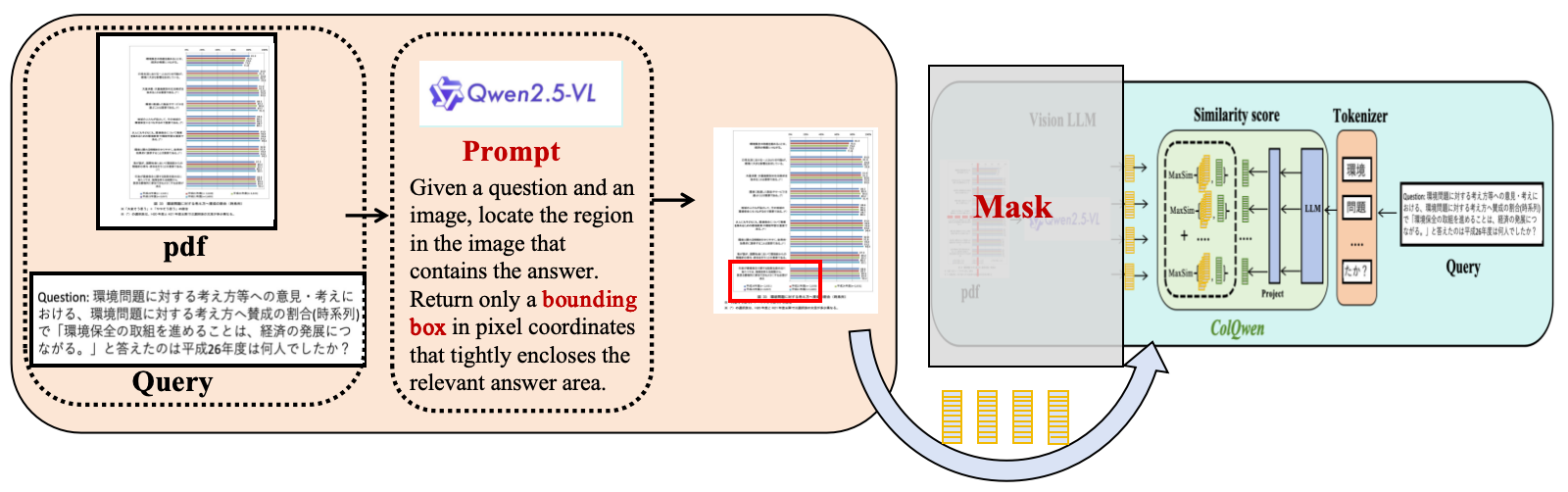}
    \caption{Extra module for page image retrieval.}
    \label{fig:extra_module}
\end{figure}
As shown in Figure~\ref{fig:extra_module}, the question and the complete image are first fed into MLLMs, exemplified by QwenVL2.5. 
A carefully crafted prompt is used to instruct the model as follows:  
\begin{itemize}
\item \textbf{First:} Analyze the provided chart or image to identify the minimum bounding rectangle that encloses all visual elements required to answer the given question.
\item \textbf{Second:} Please generate a structured JSON output containing: a step-by-step analysis of the relevance of each visual element; the normalized coordinates $[x_1, y_1, x_2, y_2] \in [0,1]^4$ of the answer-critical region; and a confidence score $\alpha \in [0,1]$ indicating localization certainty (1 = maximum confidence). 
\end{itemize}
This prompt guides the model to precisely locate the image region most relevant to answering the question.
Based on these coordinates, the original image is cropped to retain only the answer-relevant portion. 
Subsequently, the similarity between the cropped image and each of the initially retrieved top-3 images is computed using ColQwen. 
If the similarity score of the cropped image exceeds that of any original top-3 image, the cropped image replaces the corresponding original image in the retrieval set, thereby improving the semantic alignment between the retrieved images and the given question.

\section{Experiments}

\subsection{Experiment Setup}
We perform zero-shot inference directly on the test samples, and the results are obtained based on a voting mechanism: First, the model conducts three independent reasoning trials per question, shuffling the order of the options each time. The answer receiving at least two votes is selected; second, if there is a tie (e.g., each answer receives one vote), the system initiates a second-round query while retaining all candidate options; finally, the final decision is made through one more round of inference.

\begin{table*}[!h]
\centering
\caption{Performance comparison of various MLLMs on the challenge public dataset. The highest-scoring models are highlighted in red.}
\label{tab:vlm_comparison}
\resizebox{\textwidth}{!}{ 
\begin{tabular}{ccccccccccccc|c}
\toprule
\rotatebox{45}{\textbf{Model}} & \rotatebox{45}{GPT-4o} & \rotatebox{45}{Gemini-1.5} & \rotatebox{45}{Claude-3.5} & \rotatebox{45}{Llama-3.2-V-90B-I} & \rotatebox{45}{InternVL3-8B} & \rotatebox{45}{InternVL3-9B} & \rotatebox{45}{InternVL3-78B} & \rotatebox{45}{Qwen2.5-VL-7B} & \rotatebox{45}{Kimi-VL-A3B} & \rotatebox{45}{GLM-4.1-9B} & \rotatebox{45}{Qwen2.5-VL-72B-I} & \rotatebox{45}{Qwen2.5-VL-72B} & \\
\cmidrule{1-13}
\textbf{Open Source} & \xmark & \xmark & \xmark & \cmark & \cmark & \cmark & \cmark & \cmark & \cmark & \cmark & \cmark & \cmark &\\ \cmidrule{1-13}
\textbf{Base} & - & - & - & Llama-3 & \shortstack{InternViT-6B\\+Qwen2.5} & \shortstack{InternViT-300M\\+InternLM3} & \shortstack{InternViT-6B\\+Qwen2.5} & \shortstack{ViT-675M\\+Qwen2.5} & \shortstack{ViT-Base\\+Moonlight} & \shortstack{ViT-675M\\+GLM-4.1} & \shortstack{ViT-675M\\+Qwen2.5} & \shortstack{ViT-675M\\+Qwen2.5} &  \\ \cmidrule{1-14}
\textbf{Single} & 0.52 & 0.53 & 0.53 & 0.53 & 0.50 & 0.49 & 0.54 & 0.56 & 0.56 & 0.55 & 0.57 & \textcolor{red}{\textbf{0.57}} & \textbf{Score} \\ \cmidrule{1-14}
\multirow{4}{*}{\textbf{Fusion}} & & & &  & \cmark  & \cmark  & \cmark  &  & &  &  &  &0.56 \\ 
& & & &  &  & &  &  & \cmark & &  & \cmark  &0.58 \\
& & & &  &  & &  &  & \cmark & & \cmark  & \cmark  &0.58 \\
& & & &  &  & &  &  &\cmark   & \cmark  & & \cmark  & \textcolor{red}{\textbf{0.59}}\\ 
\bottomrule
\end{tabular}
}
\end{table*}

\subsection{Overall Result and Analyse}

Table~\ref{tab:vlm_comparison} presents the performance comparison of single MLLMs and fusion models on the challenge dataset, measured by the Public Score.

The results reveal two important patterns regarding the capabilities of individual models. First, the performance of Qwen2.5-VL-72B-Instruct is comparable to its base version Qwen2.5-VL-72B, indicating that instruction tuning does not bring significant improvement for this task. Notably, some smaller-scale models perform better; for example, Qwen2.5-VL-7B-Instruct and Kimi-VL-A3B outperform larger models such as InternVL3-78B and Llama-3.2-90B-Vision-Instruct, suggesting that model architecture and training data quality are more important than parameter scale alone.
Regarding fusion strategies, three empirical observations emerge. The baseline improvement shows that pure InternVL3 series model ensembles elevate the best single-model performance from scoring 54 to 56. The optimal experimental ensemble, combining Qwen2.5-VL-72B, Kimi-VL-A3B-Thinking, and GLM-4.1-9B-Thinking models, achieves the highest performance of scoring 59 . However, we observe marginal effects in model ensembles. For example, the ensemble of Qwen2.5-VL-72B and Kimi-VL-A3B-Thinking reaches scoring 58, and further adding Qwen2.5-VL-72B-Instruct maintains the score at 58, reflecting a threshold effect in model complementarity.

\subsection{Ablation Experiment}
\begin{figure}[!h]
    \centering
    \includegraphics[width=0.4\textwidth]{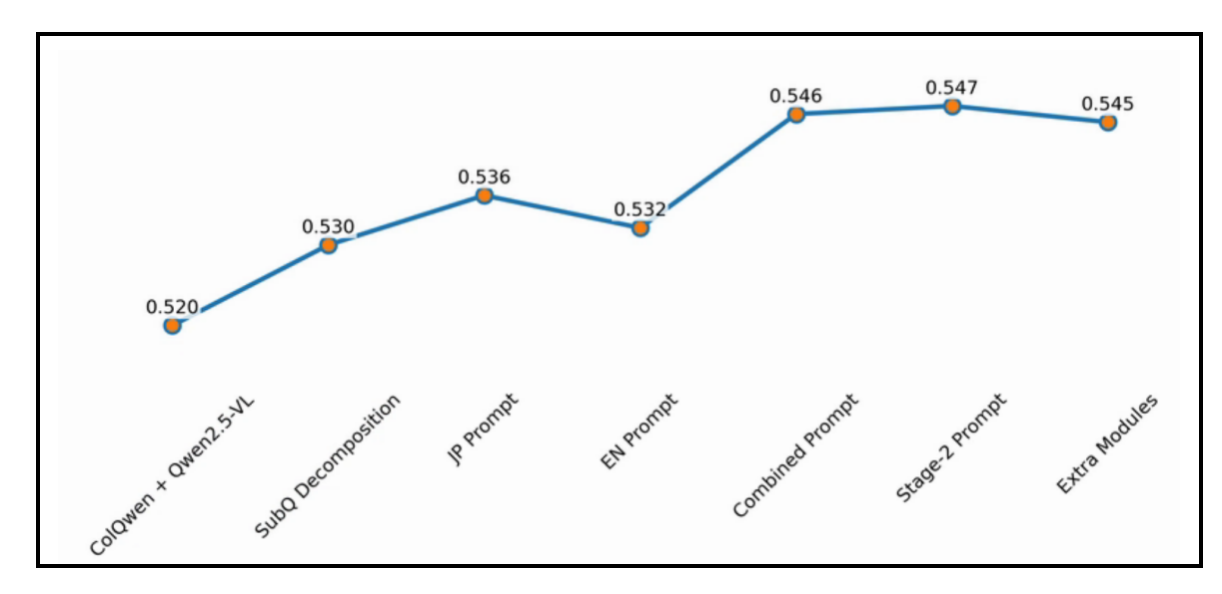}
    \caption{Ablation study on GLM-4.1: Impact of individual components on zero-shot reasoning performance.}
\label{fig:Ablation}
\end{figure}

As shown in Figure~\ref{fig:Ablation}, the baseline with only ColQwen retrieval and Qwen2.5-VL filtering scores 0.52.
Sub-question decomposition improves performance to 0.53, and first-stage multilingual prompts further raise it to 0.54, indicating that combining English and Japanese mitigates language-specific biases.
Second-stage prompt refinement and additional modules maintain this peak, suggesting that performance gains primarily arise from the synergy between retrieval filtering, decomposition, and multilingual prompt design.

\subsection{Discussion}
Firstly, although the competition training set provides rich, high-quality annotations with clear domain priors, we deliberately avoid instruction tuning and instead adopt a zero-shot inference strategy, keeping model parameters unchanged and relying solely on the general knowledge acquired during pre-training.  
We also attempted instruction tuning for this task, but performance was lower than that of zero-shot inference. Possible reasons include the high task difficulty (a large number of candidate answers), limited fine-tuning data, and the mismatch between competition questions—often based on entire PDF documents—and the PNG images used for training, which makes it challenging for the model to learn the fine-grained feature patterns required for the task.

\begin{figure}[!h]
    \centering
    \includegraphics[width=0.4\textwidth]{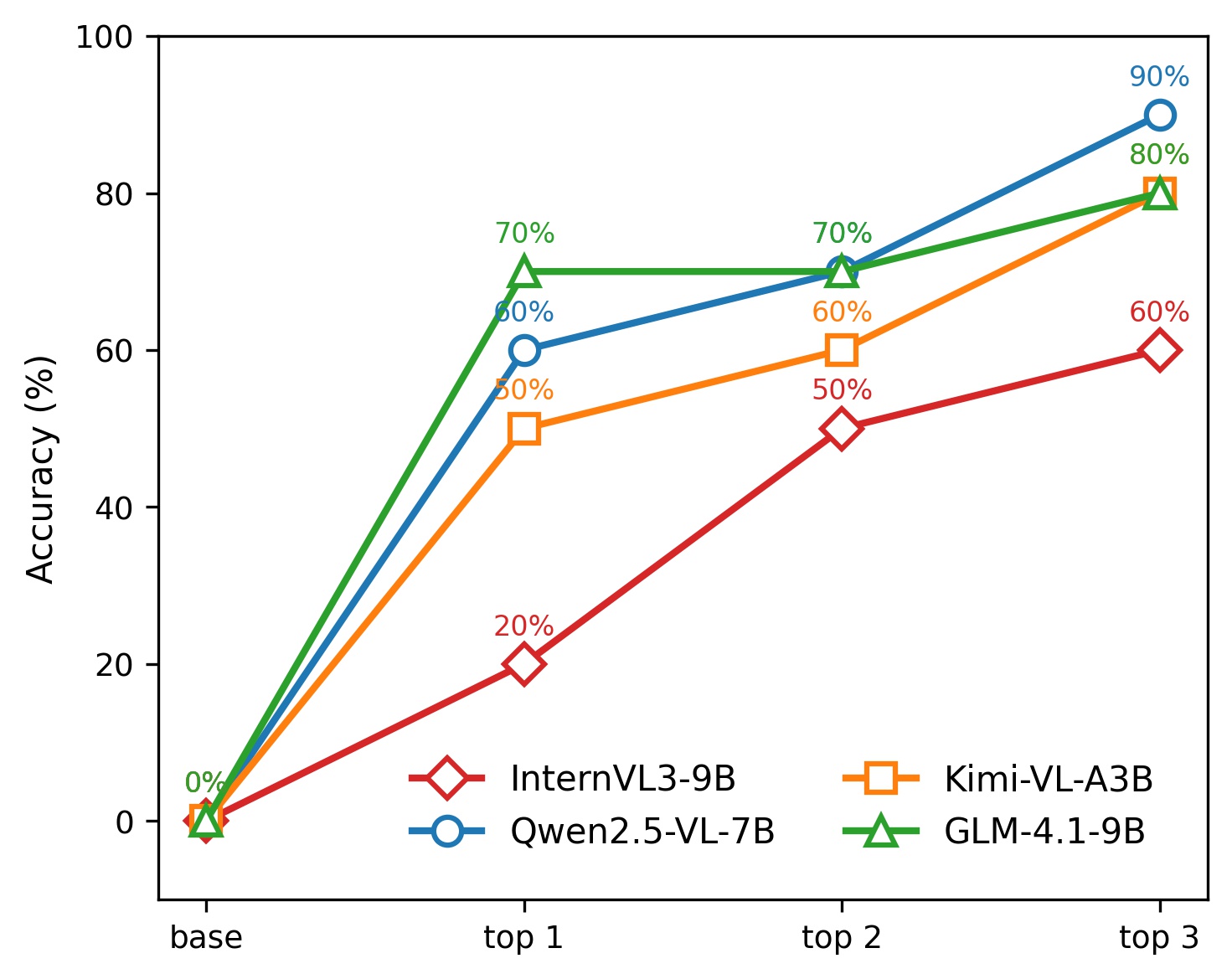}
    \caption{Case study using manually given area coordinates on 10 wrong samples.}
\label{fig:case_study}
\end{figure}
As illustrated in Figure~\ref{fig:case_study}, our detailed analysis of 10 representative failure cases identifies retrieval precision as the primary performance bottleneck. Experimental results demonstrate that when expanding the number of human-annotated key regions from one to three, the system accuracy shows substantial improvement from 50\% to 80\%--90\%. This significant performance improvement reveals a critical discrepancy: while existing models demonstrate competent reasoning capabilities when provided with relevant visual information, current automated retrieval systems still exhibit substantial deficiencies in autonomously locating these crucial regions.

\section{Conclusion}
In this study, we address the challenge of Japanese PDF document understanding by designing a framework based on a newly established benchmark dataset, which integrates multimodal information processing with hierarchical reasoning. Our approach effectively handles document layout complexity and long-range dependencies through precise image-text retrieval using ColQwen and a sub-question based semantic verification mechanism. The proposed framework decomposes complex queries and generates structured reasoning processes. Experimental results demonstrate the effectiveness and interpretability of our framework in complex Visual Question Answering tasks.


\section{Acknowledgments}
This work is supported by the National Natural Science Foundation of China under Grants Nos. 62441225, 61972192, 62172208, 61906085.
This work is partially supported by Collaborative Innovation Center of Novel Software Technology and Industrialization.
This work is supported by the Fundamental Research Funds for the Central Universities under Grant No. 14380001.
\bibliographystyle{ACM-Reference-Format}

\end{document}